\documentclass[aps,prl,twocolumn]{revtex4}

\pdfoutput=1
\usepackage{amsmath,amsfonts,amssymb,amsbsy,amscd,amsgen}
\usepackage{subfigure}
\usepackage{float}
\usepackage{float}
\usepackage{color,graphicx}
\def\V2D#1{\vec{#1}} 
\def\VMD#1{\underline{#1}} 
\def\MM#1{\underline{\underline{#1}}} 
\newcommand{\grad}{\boldsymbol{\nabla}}

\newcommand{\red}{}

\begin{document}

\title{Algorithm for a microfluidic assembly line}
\author{Tobias M. Schneider}
\affiliation{School of Engineering and
Applied Sciences, Harvard University, 29 Oxford Street, Cambridge, MA 02138}
\author{Shreyas Mandre}
\affiliation{School of Engineering and
Applied Sciences, Harvard University, 29 Oxford Street, Cambridge, MA 02138}
\author{Michael P. Brenner}
\affiliation{School of Engineering and
Applied Sciences, Harvard University, 29 Oxford Street, Cambridge, MA 02138}

\begin{abstract}
Microfluidic technology has revolutionized the control of flows at small scales giving rise to new possibilities for assembling complex structures on the microscale. We analyze different possible algorithms for assembling arbitrary structures, and demonstrate that  a sequential assembly algorithm  can manufacture arbitrary 3D structures from identical constituents. We illustrate the algorithm by showing that a modified Hele-Shaw cell with 7 controlled flowrates can be designed to construct the entire English alphabet from particles that irreversibly stick to each other.
\end{abstract}

\maketitle

Developing novel methods for assembling complex structures from small particles has been the focus of much recent investigation 
\cite{Whitesides2002,Assembly}.
Traditional approaches 
have
mostly revolved around designing selective interactions between constituent elements of the assembly. The structure is then assembled in the absence of any external control if thermal fluctuations can drive the
system to its energetic ground state \cite{Whitesides2002,MinEnergy},
 although non-trivial energy landscapes often render this approach challenging.

Here we consider a different possibility for assembly on small scales, in which microfluidic flow control is used to steer and assemble
small particles into structures of high complexity. The basic idea
follows from the
observation that if we could construct an {\sl arbitrary} time dependent flow field $\V2D{v}(\V2D{x},t)$, then particles in the flow could be advected along arbitrary paths and moved to arbitrary locations at a fixed time.
This apparently allows to construct any complex structure, with the individual components binding irreversibly upon contact.

\begin{figure}
\vspace{-12 pt}
\[
\includegraphics[width=0.45\textwidth]{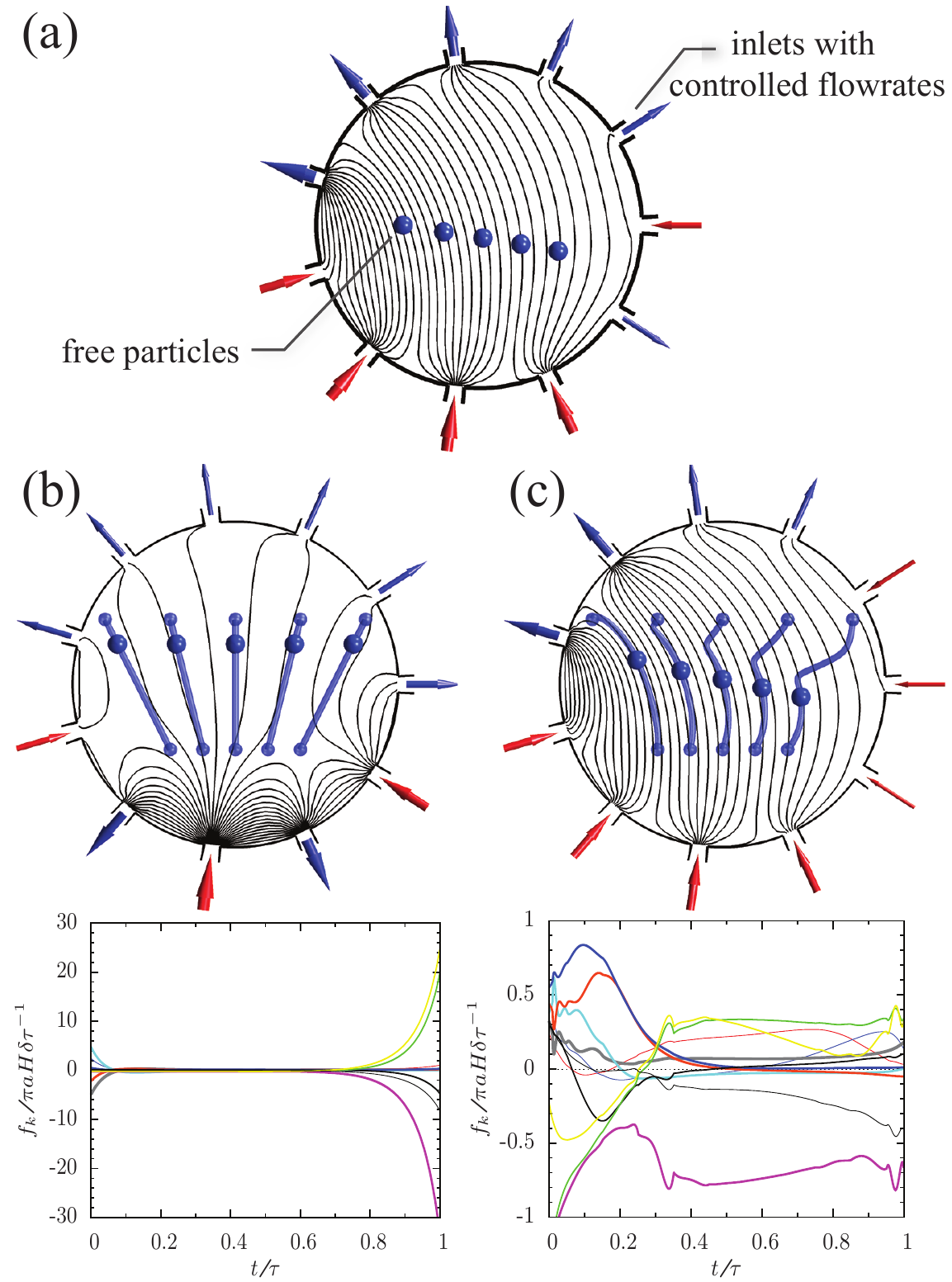}
\]
\vspace{-32 pt}
\caption{\label{fig:FigOne}
(Color online) (a) Schematic: 
Five particles are advected by the flow field in a circular
cell.
The flow field, visualized by its streamlines, is set up by eleven flowrates imposed on the boundary (arrows indicating the strength and direction).
(b/c) Two 
trajectories transporting particles from a fixed initial to their desired final position 
with the required flowrates. The linear trajectory (b) requires flowrates almost 2 orders of magnitude higher than the trajectory in (c). This optimized trajectory minimizes dissipation. 
Flowrates are given in units of the rate required to transport a single particle in one time unit across the cell. 
}
\end{figure}

Of course, the flow field $\V2D{v}(\V2D{x},t)$ cannot be arbitrary; it must conserve mass and momentum \cite{HappelBrenner},
\begin{equation}
\grad\cdot\V2D{v} = 0, \quad- \nabla p + \mu \nabla^2 \V2D{v} +  \rho \V2D{b}= 0.
\label{eqn:stokes}
\end{equation}
where $p$ is the pressure, $\mu$ the liquid viscosity and $\V2D{b}$ is a volumetric force. The question of what structures can be built thus hinges on  what flow fields can be produced using current technology \cite{Technology}, 
and what are the limits for the possible structures that can form with such flow fields.  Volumetric forces can be produced using e.g. magnetic fields 
\cite{MagFields} 
or optical tweezers 
\cite{Grier}; 
alternatively, the flow can be generated by inlets specifying  $\V2D{v}$  at the boundary of a cell.

We analyze the case where pressure inlets around the boundary force the flow (Fig.~\ref{fig:FigOne}), and $\V2D{b}=0$.  Fluid mechanical constraints prohibit simultaneous control of  many particles with this device; however, we demonstrate that a sequential assembly algorithm allows the assembly of arbitrary structures. The device can be designed to manufacture the entire English alphabet using 7 controlled flow rates in two dimensions.

\paragraph{Linear Response to Boundary Forcing.}
Consider $N$ particles suspended in the flow domain $\Omega$ with their instantaneous positions being $\V2D{x}_j$, $j=1,2,\dots, N$. Let the flow be forced on the boundary of the flow cell by a prescribed velocity $\V2D{v}(\V2D{x}, t)$, $\V2D{x} \in \partial\Omega$.
The linearity of \eqref{eqn:stokes} implies the velocity of the suspended particles is linear in the 
boundary forcing \cite{HappelBrenner}, i.e.
\begin{eqnarray}
 \sum_{k=1}^{N} R_{jk} \dot{\V2D{x}}_k = \int_{\partial\Omega} K_j(\V2D{x}_1, \V2D{x_2}, \dots \V2D{x_N}; \V2D{\xi} )~\V2D{v}(\V2D{\xi}, t)~dS 
+ \V2D{F}_j,
 \label{eqn:bem}
\end{eqnarray}
where $\V2D{F}_j$ is the force acting on the $j^\text{th}$ particle, possibly due to non-hydrodynamic interparticle interactions. The response coefficients 
$K_j$ and $R_{jk}$ depend on the geometry of the flow cell \cite{Pozrikidis2006} and can be computed numerically.
If, the boundary forcing occurs through $M$ discrete inlets of area $\Delta S$, located at $\V2D{\xi_k}$ with prescribed velocities $\V2D{v}_k$, $k=1,2,\dots, M$, 
then Eqn. \eqref{eqn:bem} can be written in the discrete form
\begin{equation}
\MM{R}(\VMD{x})~\dot{\VMD{x}} = \MM{M}(\VMD{x}) \cdot \VMD{f} + \VMD{F},
\label{eq:linear}
\end{equation}
with $\VMD{x} = [\V2D{x}_1, \V2D{x}_2, \dots, \V2D{x}_N]$, flowrates $\VMD{f}=[\V2D{v}_1, \V2D{v}_2,\dots, \V2D{v}_M]\Delta S$ 
and $\MM{M}_{jk} = K_j(\V2D{x}_1, \V2D{x_2}, \dots \V2D{x_N}; \V2D{\xi}_k )$.

Eqn. \eqref{eq:linear} is  
an instantaneous linear relation between the imposed flowrates and the particle velocities. This implies that  {\sl prescribed} particle trajectories $\VMD{x}(t)$, may be realized by imposing suitable $\VMD{f}$ 
obtained by inverting \eqref{eq:linear}.  The feasibility of this method then hinges crucially upon invertibility of $\MM{M}$, 
which in general needs not be square.

\paragraph{Condition for Controlling Particle Trajectories.}
A necessary condition for inverting $\MM{M}$ is that the number of independent controls exceed the number of degrees of freedom. With N particles in 3 dimensions,   there must be at least $M=3N+1$ flow inlets;  $3N$ inlets control particle degrees of freedom, with the additional inlet enforcing volume conservation. Similarly in two dimensions at least $2N+1$ flow inlets are required.

In general, these algebraic conditions are not sufficient for the practicality of this assembly method. We will see below that in practice, the flowrates required to independently steer large number of particles can be too large to be practical, owing to the poor conditioning of  the matrix
 $\MM{M}$.

\paragraph{Hele-Shaw Flow as a Specific Example.}
To explore this in more detail, we consider a specific example.
\red{We consider the fluid mechanics of typical microfluidic devices \cite{Devices} where the vertical gap thickness $H$ is much smaller than the lateral extent. In Hele-Shaw approximation \cite{Pozrikidis2006}, the velocity profile across the gap is a parabola, with the gap-averaged velocity $\V2D{u}$ being proportional to the gradient of the pressure field $p$ satisfying Laplace's equation. A particle 
at position $\V2D{x}_k$ responds to the flow by moving at a speed proportional to the local fluid velocity, $\dot{\V2D{x}}_k = \beta \V2D{u}(\V2D{x}_k)$, with $\beta$ depending on the particle size and shape \cite{Pozrikidis2006}. 
This linear dependence of the particle velocity on the gap-averaged fluid velocity is a general consequence of the linearity of Stokes flow. The quantitative value for $\beta$ can be calculated in two different limits:
If the particle is much smaller than the gap, it is advected by the local velocity,
so 
$\beta$ depends on the location of the particle relative to the walls. 
Alternatively, if the particle approximately spans the gap width, the separation of scale underlying the Hele-Shaw approximation breaks down near the particle. The factor $\beta$ can now be calculated by solving the Stokes equations close to the particle, matching the solution with the parabolic Hele-Shaw flow in the far field.}

We  consider the device depicted in Fig. 1(a),  a circular domain of radius $a$ with flowrates $f_k$ prescribed  at the boundary at $M$ discrete inlets
 with positions $\V2D{R}_k$.
The velocity field at any position $\V2D{x}$ 
is then given by
\begin{equation}
\V2D{u}(\V2D{x}) = - \frac 1 {\pi H}\sum_{k=1}^M \frac {\V2D{x} - \V2D{R}_k} {(\V2D{x} - \V2D{R}_k)^2} f_k \equiv \MM{B}(\V2D{x}) \cdot \VMD{f}.
\end{equation}
 Thus, the matrix $\MM{M}$ in Eqn. \eqref{eq:linear} may be constructed by combining the $N$ position dependent matrices $\MM{B}(\V2D{x_i})$  corresponding to each particle.

We first test whether this flow cell will allow simultaneous control of all $N$ particles by boundary inlets.
The flow rates scale inversely with the duration of the assembly so that the scale for the flux depends on the chosen timescale. We scale flowrate by the flux required to move a single particle across the cell  $F=\pi aH\delta/\tau$, where $\delta$ is the width of the inlet and the duration $\tau$ of the process. 
Fig.~\ref{fig:FigOne}(b) shows that if we require the particles to be transported in straight lines at constant velocity from their initial position to their final position, the required flowrates reach
up to 30 times this value. This happens because when two or more particles are moving towards each other a distance $\epsilon$ apart, a strain rate field of $\dot\epsilon/\epsilon$ is required. If the approach velocity is constant, the strain diverges as $\epsilon\to 0$, implying diverging flowrates.  Thus, whenever particles are brought together at constant velocity, large fluxes are required.

\paragraph{Optimized Trajectories.}
We can try to reduce the flow rates  by  choosing the particle trajectories and speeds connecting the initial and the final states to minimize this effect. Intuitively, we can decrease the speed of the particles as they approach each other to minimize the required flowrates. To find out if this is sufficient
we compute the optimal trajectories connecting the initial to the desired particle configuration by finding the trajectories that minimize  dissipation.  The dissipation rate is given by
\begin{equation}
w = \frac{H^2}{12 \mu}\int_A |\nabla p|^2 dA.
\end{equation}
For the discretized forcing $w$ can be expressed as a quadratic form in the flowrates $w=\VMD{f^\dag}\cdot\MM{D}\cdot \VMD{f}$ with metric
 $\MM{D}$ being analytically given in terms of the inlet positions.
We minimize the dissipation and thereby the flowrates under the constraints that the dynamics move the particles according to their equation of motion Eqn.~\eqref{eq:linear} from their chosen initial state to their final state. This requires that we find the trajectories that minimize the Lagrangian \begin{equation}
\mathcal{L} = \int_0^1 dt \left\{\VMD{f^\dag}\cdot\MM{D} \cdot \VMD{f}
-\VMD{\lambda}^{\dag} \cdot \left[ \dot{\VMD{x}} - \MM{M}\cdot \VMD{f}\right]
- \gamma \;\VMD{e}^{\dag}\cdot\VMD{f}
\right\}
\end{equation}
where $\VMD{\lambda}$ and $\gamma$ are Lagrange multipliers to enforce the constraints and $\VMD{e}^{\dag} =(1,\ldots,1)$.    $\lambda$ enforces 
the equation of motion, whereas $\gamma$ requires that the fluxes $\VMD{f}$ satisfy volume conservation.

To minimize $\mathcal{L}$ we consider a small perturbation  of the flowrates  in the direction of steepest descent, i.e. $\VMD{f}\rightarrow \VMD{f}+\VMD{\Delta f}$ with
\begin{equation}
\VMD{\Delta f} = - \epsilon \left( \frac{\delta \mathcal{L}}{\delta \VMD{f}}\right)^{\dag}=- \epsilon \left\{ 2 \MM{D}\cdot \VMD{f}+ \MM{M}^{\dag}\cdot\VMD{\lambda}-\gamma \VMD{e}\right\}.\label{eq:fluxupdate}
\end{equation}

To fix the unknown fields $\gamma$ and $ \VMD{\lambda}$ we require the variation of $\mathcal{L}$ with respect to $\VMD{x}, \VMD{\lambda}$ and $\gamma$ to vanish for the updated flowrates and trajectory 
 $\VMD{x}\rightarrow \VMD{x}+\VMD{\Delta x}$.
Evaluating the stationarity condition at linear order in the increments
allows to eliminate $\gamma$ from Eqn.~\eqref{eq:fluxupdate} and selects a unique solution of the \emph{adjoint equations}  
\begin{equation}
\dot{\lambda}_k = -f_n \frac{\partial M_{mn}}{\partial x_k}\lambda_m
\end{equation}
which are numerically solved together with the evolution of $\VMD{x}$ using MATLAB.
\footnote{Since the trajectory increment 
preserves constraints to linear order only, we explicitly enforce in each iteration that particles reach their prescribed final positions exactly. This corrects for neglected higher order effects.}

The minimization starts from the linear trajectory in Fig.~\ref{fig:FigOne}(b) along which we choose velocity variations damping  
large peak flowrates. Panel (c) shows the same 5 particles following an optimized trajectory
for which the flowrates are reduced by more than an order of magnitude, to the desired range. The reduction 
is due to choosing $\dot\epsilon/\epsilon$ to approach a finite number as $\epsilon \to 0$.  Thus, optimizing trajectories can lead to simultaneous control of a modest number of particles using boundary flowrates.

Nonetheless, this approach does not scale to larger number of particles;
e.g,
we could not find trajectories allowing  simultaneous control of 13 particles
required to spell the letter `B' at moderate fluxes. This reflects a 
implementation independent 
physical limit of directing flow fields with boundary fluxes, resulting from the fact that flow modes forced at the boundary  decay in amplitude as one moves away from the wall.  The characteristic length scale governing the decay of the boundary modes is the distance between the injection points which scales inversely with the number of inlets. But as the particle number increases, we need more inlets to control the particle motion.  This 
limits the number of particles that 
can be controlled simultaneously.  In our numerical experiments we could not find trajectories transporting more than 6 individual particles at moderate flowrates.

\paragraph{Sequential Assembly.}

To overcome this physical limitation,  an algorithm is required that decouples the number of controlled degrees of freedom from the number of particles in the desired structure. This can be achieved with a sequential approach, where one particle after the other is attached to an aggregate which moves as a rigid body subject to force- and torque balance. Construction of the corresponding $\MM{M}$ matrix requires explicitly accounting for the translation and rotation of the cluster consisting of $Z$ particles at positions $\V2D{x}_\alpha = \V2D{x}_{cm} + \MM{R}(\varphi) \left( \V2D{\xi}_\alpha - \V2D{\xi}_{cm}\right)$
where $\V2D{\xi}_\alpha$ are the prescribed particle positions in the aggregate's frame of reference and $\MM{R}$ is the two-dimensional rotation.
How many degrees of freedom are required for the assembly? 
Controlling the position of the aggregate requires two degrees of freedom. Rotating it requires a stagnation point flow superimposed on the local mean flow near the aggregate. This requires independent control of two more degrees of freedom, the strength and orientation of the stagnation point flow. Finally, the free particle location demands two additional degrees of freedom. Including volume conservation we thus need 7 inlets to absolutely control all degrees of freedom. 
\begin{figure}[b]
\vspace{-27 pt}
\[
\includegraphics[width=0.45\textwidth]{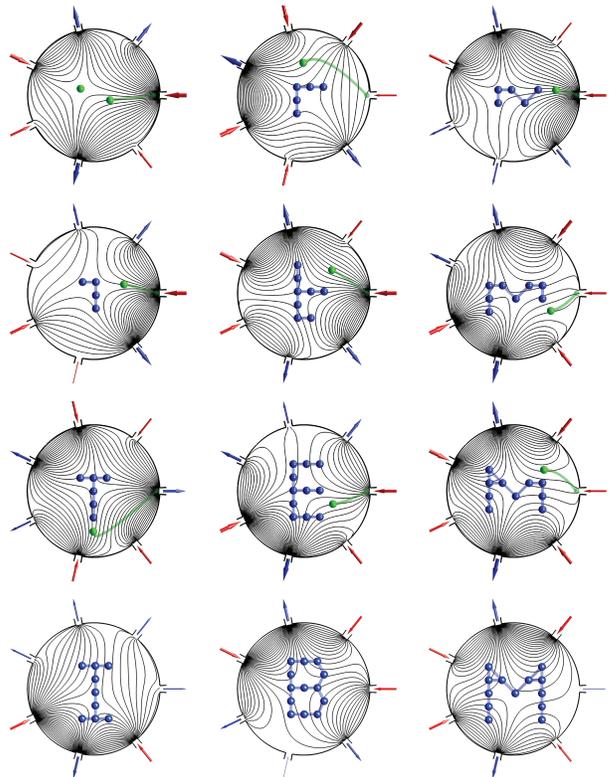}
 \]
\vspace{-30 pt}
\caption{\label{fig:SpellIBM}
(Color online) Sequentially particles are added to form aggregates of arbitrary shape. The three columns present snapshots along the assembly of three example structures presented in the last row. The required time-variation of the seven controlled flowrates are computed a priori (eg. Fig.~\ref{fig:Bflux}). See movie M1 (online) for an animation. }
\end{figure}

\begin{figure}
\vspace{-15 pt}
\[
\includegraphics[width=0.43\textwidth]{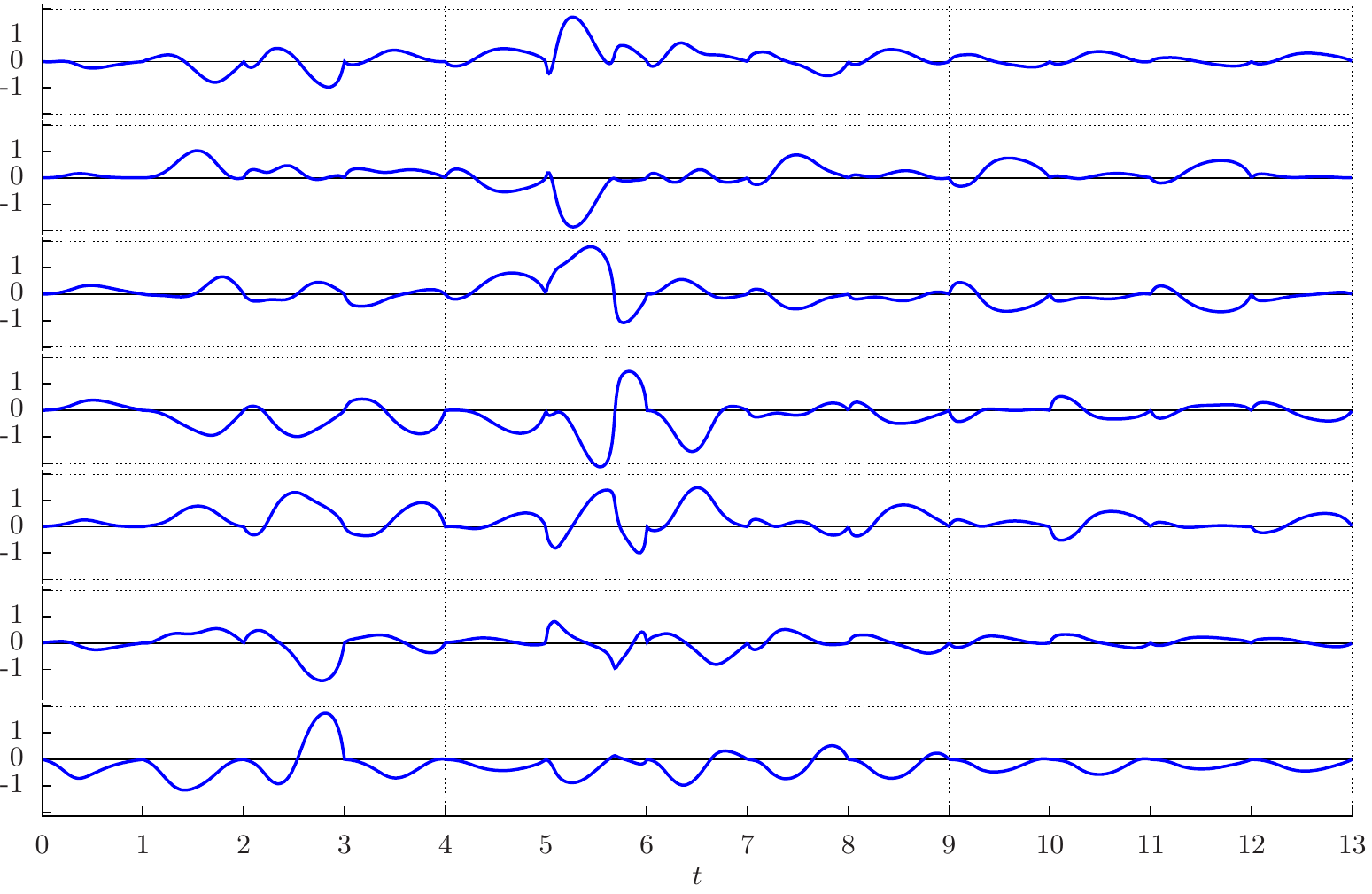}
\]
\vspace{-32 pt}
\caption{
(Color online) 
Flowrates $f_k / \pi aH\delta\tau^{-1}$ for the seven inlets as a function of time $t/\tau$. These allow to spell the letter `B'. At each integer time 
a new particles enters the cell. \label{fig:Bflux}
}
\end{figure}

Sequentially adding particles 
allows to assemble structures of arbitrary shape and thus spell a word (Fig.~\ref{fig:SpellIBM})
without any feedback control.
Technically, the trajectory for a particle to be added is constructed by defining the desired position which is taken from \cite{Eigler1990} and the direction of approach in the frame of reference of the aggregate.
Spline interpolation between the initial position of the new particle and its final position 
when the aggregate has reached the desired configuration, then yields the particle path. The aggregate can also be moved arbitrarily; we choose to keep its orientation fixed and move the center of mass to the device center in each step.
The velocity along this trajectory is chosen such that it vanishes at the initial and final configuration so that flowrates reach zero after each assembly step.
With this choice of trajectories, the  flowrates are computed inverting Eqn.~\eqref{eq:linear}.
No extensive optimization was required to limit the fluxes (see Fig.~\ref{fig:Bflux}) to reasonable values of the same order as those in Fig.~\ref{fig:FigOne}(c). Optimizing trajectories might however still be useful either to further reduce the required flowrates or to minimize internal forces within the growing aggregate so that the forming structure can sustain the stresses in the flow.

We envision a typical mode of operation of the microfluidic device similar to that of a macroscopic robotic assembly line: Once the constituents of an assembly and the detailed assembly sequence are decided on, the flowrates on the inlets of the device and the precise instants at which each constituent is introduced are calculated. 
These pre--computed flowrates are then set up in a time--periodic fashion, and a train of the assembly constituents are introduced at the inlets of the device to accomplish a train of assembled products.

In summary, we have shown that microfluidic assembly can be an efficient strategy if structures
are built sequentially. Using this approach, we show that the temporal control of only 7 flowrates allows to build arbitrarily shaped particle aggregates in 2 dimensions. In 3 dimensions, the same argument implies that 11 different flowrates are required.
Different chamber geometries and hydrodynamic interactions can be incorporated into this framework, which rests solely on the linearity of Stokes flow.
We anticipate that similar algorithms can be constructed using other forcing mechanisms such as electrokinetics \cite{Cohen2005}, where non-hydrodynamic electrical forcing contributions need to be included into the transfer 
matrices. A challenge for practical implementation is to quantify the sensitivity of particle trajectories to noise in the imposed flow rates. This is particularly relevant when scaling down the assembly to submicron scale, where the Pecl\'et number corresponding to Brownian motion is small and hence potentially disruptive. 
One option for dealing with 
errors and noise
is to implement feedback control \cite{Cohen2006}, though this significantly complicates the process especially for three dimensional structures. Another intriguing possibility is to simply embrace the existence of stochasticity 
and construct the most stable trajectories that 
maximize the probability of formation of the desired structures in the presence of noise.  Finding algorithmic methods for carrying out this optimization are important directions for future research.

\begin{acknowledgments}
We acknowledge support by the 
National Science Foundation; 
the Kavli Institute for Bionano Science and Technology at Harvard; and the German Science Foundation through grant Schn 1167/1 (TMS).
\end{acknowledgments}


\end{document}